\documentstyle[aps,prl,twocolumn]{revtex}
\begin{document}
\title{Spinons, Solitons and Magnons in One-dimensional
Heisenberg-Ising Antiferromagnets}
\author{Rajiv R. P. Singh}
\address{Department of Physics, University of California, Davis,
California 95616}
\date{\today}
\maketitle

\begin{abstract}
We calculate the excitation spectra for
the one-$d$ Heisenberg-Ising antiferromagnets
by expansions around the Ising limit. 
For $S=1/2$, the calculated expansion coefficients for the
spinon-spectra agree term by term with the solution of Johnson
and McCoy. For $S=1$, the solitons become
gapless before the Heisenberg limit is reached, signalling
a transition to the Haldane phase. By applying a staggered
field we calculate
the one-magnon spectra for the $S=1$ 
Heisenberg chain. For $S=3/2$ the quantum renormalization
of the spin-wave spectra is calculated to be approximately $1.16$.
\end{abstract}
\pacs{PACS:}

\narrowtext

The study of spin-chains has produced many exciting ideas in
many-body physics. The combination of 
exact solutions, field theories 
and numerical studies have led to a rather
comprehensive understanding of the physical properties
of these systems \cite{review}. One of the most remarkable physics to
come out of one-dimensinal antiferromagnets is the
difference between the excitation spectra of integer
and half-integer spin-chains. The $S=1/2$ chain is integrable
and a lot is rigorously known about its excitation spectra. 
des Cloiseaux and Pearson \cite{dp} were the first to obtain
expressions for the spectrum of the Heisenberg model.
Later Johnson and McCoy \cite{jm} obtained expressions for the
low lying excitations of the Heisenberg-Ising model.

Fadeev and Takhtajan \cite{ft} argued
that the elementary excitations
for the spin-half case are spinons and have the quantum
numbers of a neutral fermion.
Using semiclassical arguments, Haldane \cite{haldane}
showed that  the topological excitations for integer
spin-chains had integer spin and those for half-integer
spin-chains had half-integer spin. He further argued that
for integer spin these solitons become gapless before
the Heisenberg limit is reached. Condensation of these
solitons leads to a novel ground state, with a gap
in the magnon excitation spectra. For half integer
spins the spectra remains similar to the spin-half case,
becoming gapless and linear at the Heisenberg point.
By now there are numerous numerical \cite{numerics} confirmations
of these remarkable facts.

In this paper we show that the spectra for the
topological excitations such as
spinons and solitons as well as the small spin-deviation excitations
or magnons can be calculated systematically by a
straightforward Raleigh-Schrodinger perturbation theory.
This is simply a consequence of the fact that 
in one-dimension a domain-wall between two ground states
is a localized object and has a well defined dispersion.
While the perturbation expansion gives essentially the
complete answer for systems with significant Ising
anisotropy, it is also very accurate for Heisenberg systems.
For $S=1/2$, our calculated coefficients for 
the spinon spectra agree term
by term with a direct expansion of
the closed form results of Johnson and McCoy \cite{jm}.
We also calculate the soliton excitation spectra for the
$S=1$ chain and find that it becomes gapless before
the Heisenberg model is reached. As shown by Haldane \cite{haldane}
these solitons play the same role as the
domain-walls in the $1+1$-dimensional classical Ising
model, where vanishing free-energy cost for the
domain-walls leads to a second order phase transitions.

In addition to the topological excitations we also
calculate the spectrum for magnons which are the quanta
for small spin-deviations around a given state.
For $S=1/2$ chain, a magnon is not a sharply defined state
even in the Ising
limit as its energy is degenerate with that of two spinons.
Thus it can break apart into two spinons.
For $S\ge 1$, magnons are well defined elementary excitations
near the ising limit. For $S=1$, we add a
staggered field term to the Hamiltonian, with a 
coefficient which
goes to zero at the Heisenberg point. The presence
of this term avoids the
condensation of solitons with non-zero Ising anisotropy,
and allows us to calculate the magnon
excitation spectra in the Heisenberg limit.
Even simple addition of available terms in the series
leads to a spectrum which agrees with previous numerical
estimates to within a few percent.

For $S=3/2$ we are not aware of any previous numerical
calculation of the spin-wave excitation spectra.
Our calculations provide extremely accurate estimates
of the spectra for systems with significant Ising anisotropy.
We find that the series also
show good convergence upto the Heisenberg limit especially
away from the gapless point near $q=0$, where they are consistent
with a linear spectra but converge more slowly. We estimate the
quantum renormalization of the spin-wave spectrum from the
maximum spin-wave energy as was done in recent experiments
on CsVCl$_3$ \cite{expt3by2}. 
We obtain this renormalization to be approximately $1.16$.
This is larger than the order $1/S$ spin-wave estimate 
of $1.12$ \cite{spinwave}
but still smaller than the central estimates of the
experiments which give $1.26$ \cite{expt3by2}. 
However, the experimental uncertainties are $\pm 0.2$
due to large uncertainties in the value of $J$. 

In the Ising limit all these excitations are localized,
and those localized at different points in space
are degenerate. The magnons consist of a single spin-deviation in a 
Neel state, whereas a soliton or a spinon is a domain wall 
between two degenerate ground states.
A simple way to think of the topological
excitations is in terms of the ground state
of a large odd-site ring. In the Ising limit, an $N$-site ring
has at least $N$ degenerate ground states, corresponding to the 
different locations where the neighboring spins are not antiparallel.
This broken bond can be thought of as the domain-wall.
Since, these ground-state configurations differ locally
from the ground state of an even site ring
only at the wall, ie the energy-density with respect to
the ground-state is non-zero only on one bond, we can regard
the excitation as localized there. 
As the $x-y$ coupling is turned on these degenerate
ground states evolve into a band of states with well-defined
energy momentum relation. Thinking in terms of an odd-site ring 
also explains why these domain-wall excitations have half-integer
spin for spin-half chains and integer spin for integral
spin-chains. In the Ising limit a spin-chain with $S>1/2$ has
other ground-state degeneracies, for example the spin right next
to the broken bond can be in any one of the $S^z$ states.
Most of the degeneracies are lifted with non-zero $x-y$ coupling
and only those related to symmetry survive.

The expansion coefficients for the
excitation spectra are obtained by a degenerate perturbation theory
which has recently been formulated in terms of a cluster expansion
by Gelfand \cite{gelfand}.
The magnon calculation is similar to that in higher
dimension \cite{sgelfand}, but the solitons are special to one-d
as discussed in the previous paragraph. For the 
purpose of our calculation 
we only need to focus on the vicinity of the excitation.
The boundaries can be 
considered to be infinitely far away and do not matter.
The degenerate perturbation theory, allows us to construct
an effective hamiltonian in the space of these 
localized Ising states, which can then be diagonalized 
straightforwardly by Fourier transformation. The recursive
procedure for obtaining the effective Hamiltonian has been
discussed by Gelfand \cite{gelfand}.

We consider the Heisenberg-Ising Hamiltonian:
\begin{eqnarray}
{\cal H}=J\sum_i S^z_iS^z_{i+1}
+ \lambda (S^x_iS^x_{i+1}+S^y_iS^y_{i+1}).
\end{eqnarray}
For the $S=1/2$ case it is most convenient to present
the square of the spinon dispersion, which we calculate
via the cluster expansion to order $\lambda^{12}$
\begin{eqnarray}
16\epsilon(q)^2&=4+20\lambda^2-6\lambda^4+{3\over 2}\lambda^6+
{3\over 8}\lambda^8-{9\over 128}\lambda^{12}
\nonumber\\
&+\cos{2q} (16\lambda+4\lambda^3-{1\over 4}\lambda^7
-{1\over 16}\lambda^9+{1\over 64}\lambda^{11})\nonumber\\
&+O(\lambda^{13})
\end{eqnarray}
A closed form expression for the spinon dispersion can be read
off from the work of Johnson and McCoy \cite{jm}:
\begin{eqnarray}
\epsilon(q)=(1-\lambda^2)^{1/2}{K\over\pi}(1-k^2\sin^2q)^{1/2}
\end{eqnarray}
Here $K$ ( and similarly $K^\prime$ )
is a complete elliptic integral of the first kind
with modulus $k$. The latter is determined as a function of
$\lambda$ from the relation
\begin{eqnarray}
{K^\prime (k)\over K (k)}= {\cosh^{-1}(1/\lambda)\over\pi}
\end{eqnarray}
We have verified that to the order calculated
our expansion coefficients agree term by term
with a direct expansion of the above expressions \cite{footnote1}.
In general, this expansion
shows excellent convergence. By simply adding the 
terms in Eq. 2, one obtains the dispersion for the
Heisenberg model as shown in Figure 1 and compared
with the exact Heisenberg spectrum. Except very
near $q=\pi/2$, where the spectrum becomes gapless at
the Heisenberg point, the convergence is excellent.
It is also evident that the maximum of the spectra at $q=0$
is much less sensitive to small anisotropy than the
spectrum near $q=\pi/2$.

For $S=1$, the soliton states can have $S^z=0,\pm 1$. While
all three are degenerate in the Ising limit, the $S^z=0$
soliton becomes the lowest energy one for $\lambda\ne 0$.
We have calculated the full spectrum for the solitons.
The minimum energy soliton corresponds to $q=\pi$, unlike
the spin-half case where the minimum energy is at $q=\pi/2$.
The mass gap is given by:
\begin{eqnarray}
m=&2-2\lambda -{2\over 3}\lambda^2+{2\over 3}\lambda^3
   -1.207407 \lambda^4   + 1.971667 \lambda^5  \nonumber\\
&-2.723225 \lambda^6   + 3.294886 \lambda^7 
  -3.400478 \lambda^8 \nonumber\\  
 & + 2.240766 \lambda^9 
   + 1.571335 \lambda^{10} +\ldots
\end{eqnarray}
Since we expect a $2D$ Ising critical point, with $\nu=1$,
where the
soliton becomes massless, we use simple Pade approximants
to obtain the location of this critical point.
We estimate
$\lambda_c=0.842\pm 0.002$, which compares well with previous
studies of this phase transition
using ground state properties \cite{ising}.

In order to calculate the magnon spectra in the Heisenberg
limit we consider the Hamiltonian:
\begin{eqnarray}
{\cal H}=&J\sum_i S^z_iS^z_{i+1}
+ \lambda (S^x_iS^x_{i+1}+S^y_iS^y_{i+1})\nonumber\\
&+(1-\lambda)
\sum_i (-1)^i S^z_i.
\end{eqnarray}
The addition of the last term ensures that for $\lambda\ne 1$,
only one of the two Neel states is the ground state of the
system and the other Neel state ( and the solitons) have
infinite excitation energies. Exactly at the Heisenberg
point ( $\lambda=1$ ) the staggered field term goes to zero
and one recovers the Heisenberg hamiltonian. Because we
have introduced the staggered field and doubled the unit cell,
the magnon spectra is also doubled. Since there is no
long range order in the Heisenberg limit, only
the calculated spectra between wavevectors ($\pi/2,\pi$) will survive
and the spectral weights for
those between ($0,\pi/2$) will vanish
as $\lambda\to 1$. The effective Hamiltonian,
whose fourier transform gives the excitation spectra is
given complete to order $\lambda^{10}$
in table 1. The convergence of the series for $\lambda=1$
is excellent. In Figure 2 we show simply the sum of terms
from $8$, $9$ and $10$ term series. Pade
approximants can be used to improve convergence slightly.
They lead to estimates for the energy gap of $0.42\pm 0.01 J$.
The entire spectrum is within a few percent of those calculated
earlier by exact numerical diagonalization of finite systems
\cite{numerics}.

For $S=3/2$ we have also calculated the effective hamiltonian
for the magnon excitation spectra. It is presented in
table 2. In
Figure 3 we show the estimated dispersion for $\lambda=1$ obtained 
by simply adding terms upto $4$th, $6$th and $8$th order.
Except very near $q=0$, where the dispersion should become
linear and gapless at the Heisenberg point,
we find that this simple addition converges rapidly.
By assuming a linear dispersion near $q=0$, and using
series extrapolation methods we can directly estimate
the spin-wave velocity.
However, if we assume that the spin-wave spectrum is
uniformly renormalized with respect to the classical
spectrum, the quantum renormalization of the spin-wave
spectrum can be estimated from the value of the
excitation energy at $q=\pi/2$. 
The latter measure, involving the maximum spin-wave energy,
is insensitive to anisotropy, whereas
the linear spectrum at small $q$ is clearly very sensitive
to anisotropy. Using the latter measure, we estimate
the quantum renormalization of the spectrum to
be approximately $1.16$. This is somewhat larger than the order 
$1/S$ spin-wave estimate of $1.12$ \cite{spinwave}.
Recently, this quantum renormalization
of the spin-wave spectra has also been determined
experimentally in the material CsVCl$_3$ 
from the energy of the zone-boundary magnons \cite{expt3by2}.
They obtain a spin-wave renormalization
of $1.26\pm 0.2$, the large uncertainties resulting from
the lack of accurate determination of the exchange constant $J$.

In conclusion, we have shown that topological excitations
such as spinons and solitons as well as
magnons in one-dimensional Heisenberg-Ising
antiferromagnets can be studied by a
straightforward application of Raleigh-Schrodinger 
perturbation theory. This expansion gives nearly the complete
answer for systems with significant Ising anisotropy, but
is also highly accurate for the Heisenberg model.
We have used this method here
to verify the spinon spectrum for $S=1/2$, 
to calculate the soliton spectrum for $S=1$ and show that it
becomes gapless at the Ising critical point, as well as to
calculate the magnon spectra for $S=1$ and $S=3/2$ chains.
The application of a staggered field, whose coefficient
goes to zero in the Heisenberg limit allows one to study
the magnon spectra for the Haldane-gap Heisenberg systems as well.

We would like to thank Martin Gelfand for discussions.
This work is supported in part by the National Science
Foundation under grant number DMR 93--18537.

\begin{figure}
\caption{The spinon excitation spectra for the spin-half Heisenberg
model obtained by adding terms upto order 10 (crosses), 
11 (circles) and 12 (squares) in Eq. 2.
The solid line is the exact result}
\label{s=1/2}
\end{figure}

\begin{figure}
\caption{The magnon spectra for the spin-$1$ Heisenberg
model obtained by adding terms upto order 8 (squares), 
9 (circles) and 10 (crosses) in table 1.}
\label{s=1}
\end{figure}

\begin{figure}
\caption{The spin-wave spectra for the spin-$3/2$ Heisenberg
model obtained by adding terms upto order 4 (crosses), 
6 (circles) and 8 (squares) in table 2.}
\label{s=3/2}
\end{figure}

\begin{table}
\caption{Effective Hamiltonian for the $S=1$ Heisenberg-Ising
model (Eq. 6). To get the magnon dispersion 
at wavevector $q$ one needs to sum the series over 
$r$ with a factor $\cos(q r)$.}
\begin{tabular}{cl}
$r$ & Series\\
\hline
  0 &
   $3 -\lambda  -0.2\lambda^2  -0.18\lambda^3   -0.0530357 \lambda^4
    +0.0088342\lambda^5$ \\
   &$ +0.0109158\lambda^6
   +0.0015458 \lambda^7
   +0.0021357 \lambda^8 $\\
   &$+ 0.0080557\lambda^9 
   +0.0105825\lambda^{10}$\\
  2 &
 $-{1\over 2}\lambda^2  -{1\over 4}\lambda^3  -0.1094048\lambda^4   
  -0.0644654\lambda^5 
  -0.0528568\lambda^6 $\\
 &$-0.0381666\lambda^7   -0.0176628\lambda^8
   -0.0020513\lambda^9 $\\
 &$ +0.0025792 \lambda^{10}$\\
  4 &
   $-0.019375 \lambda^4   -0.0245625\lambda^5   -0.0212035\lambda^6
   -0.0177130\lambda^7$\\
 &$ -0.0153450 \lambda^8 
  -0.0132463\lambda^9
   -0.0114761\lambda^{10}$\\
  6 &
  $-0.0016698\lambda^6   -0.0035734\lambda^7
   -0.0046995\lambda^8  $\\
  &$ -0.0052158\lambda^9 
  -0.0054470\lambda^{10}$\\
  8 &
   $-0.0001682 \lambda^8   -0.0004937\lambda^9
   -0.0008462 \lambda^{10} $\\
  10 &
  $ -0.0000195 \lambda^{10}$\\
\end{tabular}
\label{spinone}
\end{table}

\begin{table}
\caption{Effective Hamiltonian for $S=3/2$ Heisenberg
Ising model. To get the magnon dispersion 
at wavevector $q$ one needs to sum the series over 
$r$ with a factor $\cos(q r)$.}
\begin{tabular}{cl}
$r$ & Series\\
\hline
  0 &
$3  -0.825\lambda^2   -0.0529051\lambda^4   +0.0195540\lambda^6
 +0.0186877\lambda^8$\\
  2 &
$ -1.125 \lambda^2  -0.2552009\lambda^6  -0.0900226\lambda^8
   -0.0174543\lambda^8$\\
  4 &
  $ -0.0980859\lambda^4   -0.0691161\lambda^6
   -0.0454369\lambda^8$\\
  6 &
  $ -0.0190201\lambda^6   -0.0229853\lambda^8$\\
  8 &
   $-0.0043107\lambda^8$\\
\end{tabular}
\label{spin-3by2}
\end{table}


\begin{references}
\bibitem{footnote1} The Johnson and McCoy expressions \cite{jm},
involving elliptic functions can be used to generate
many more terms in the expansion, however we shall not
attempt to do so here.
\end{references}
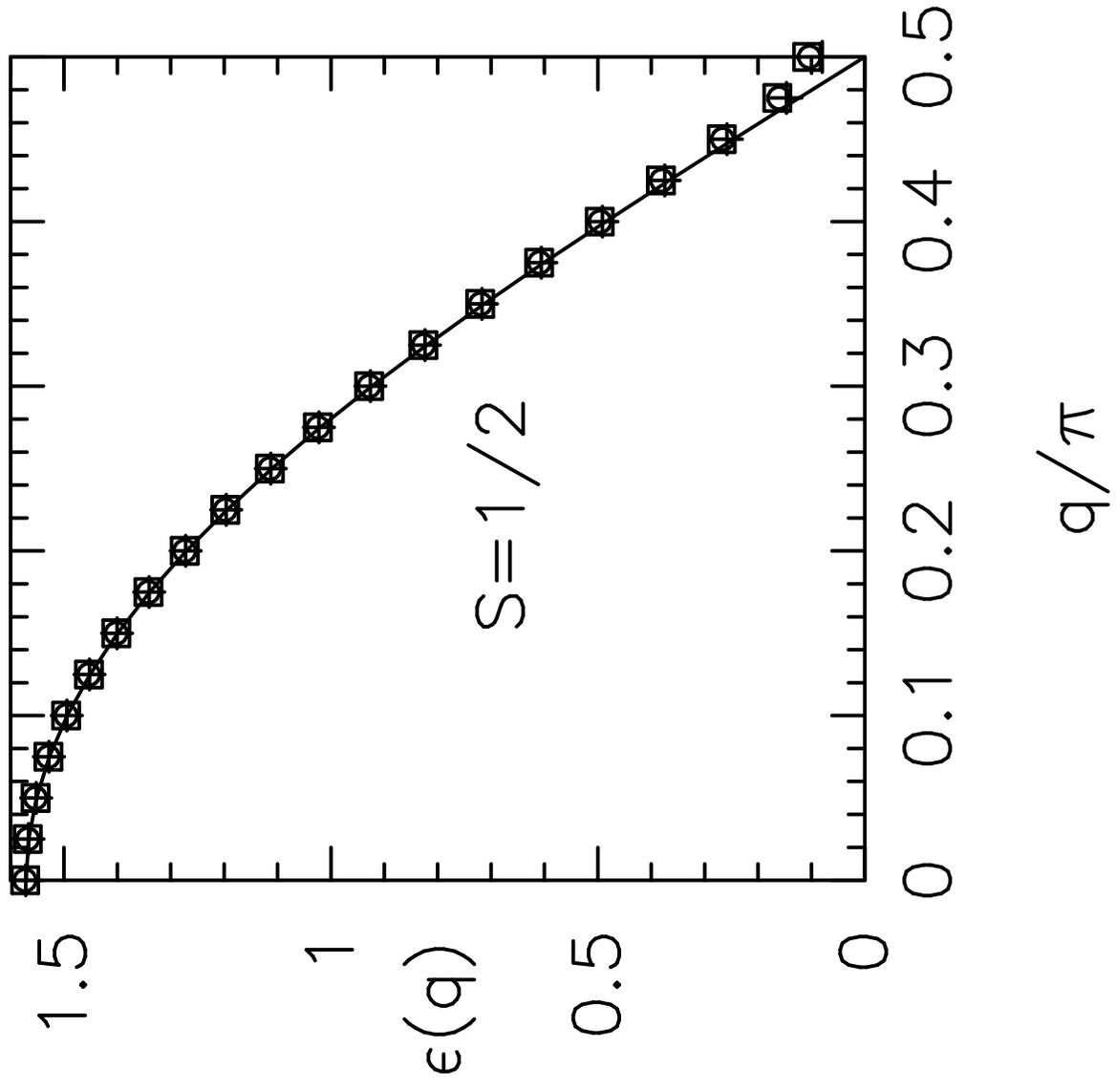
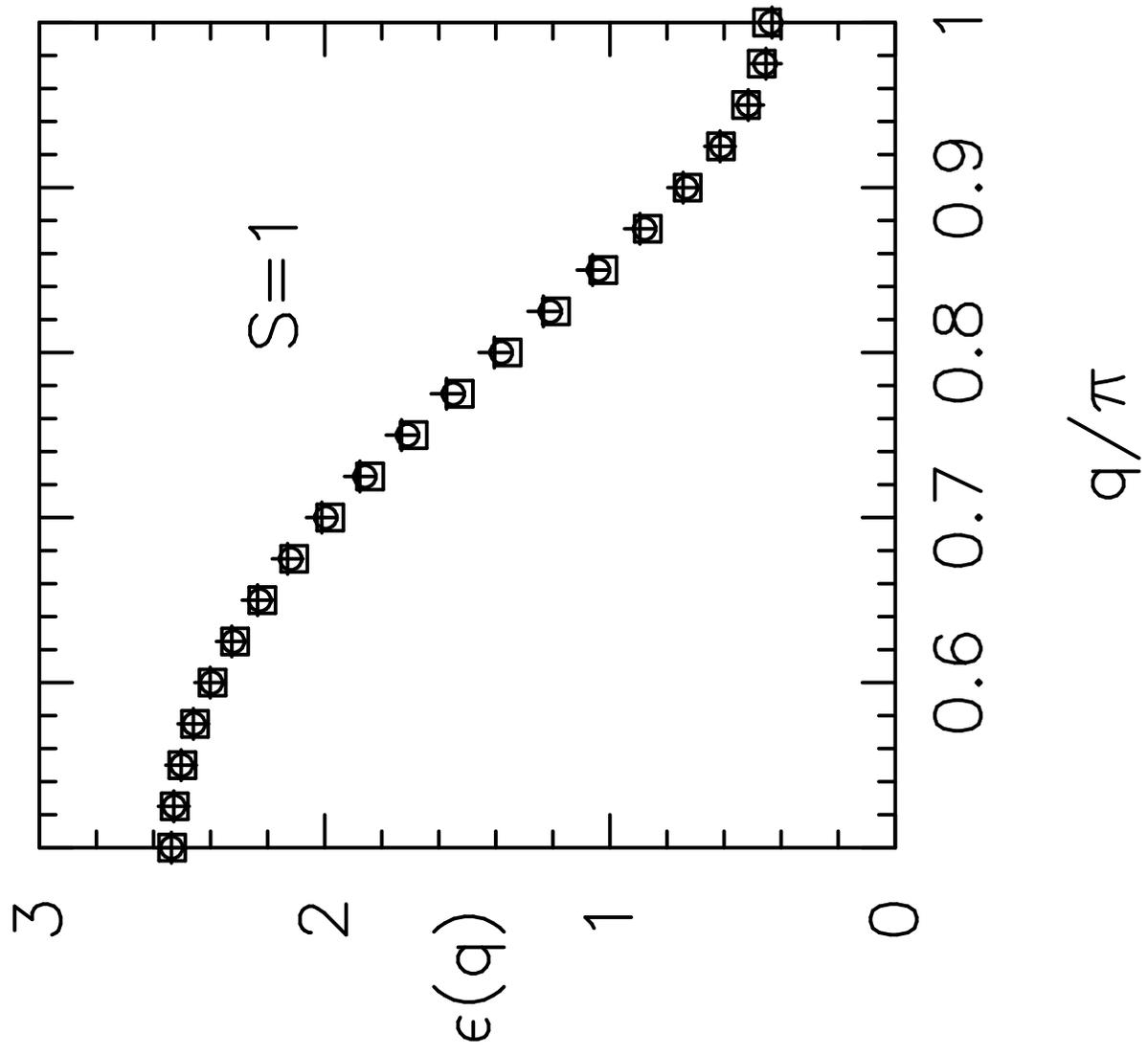
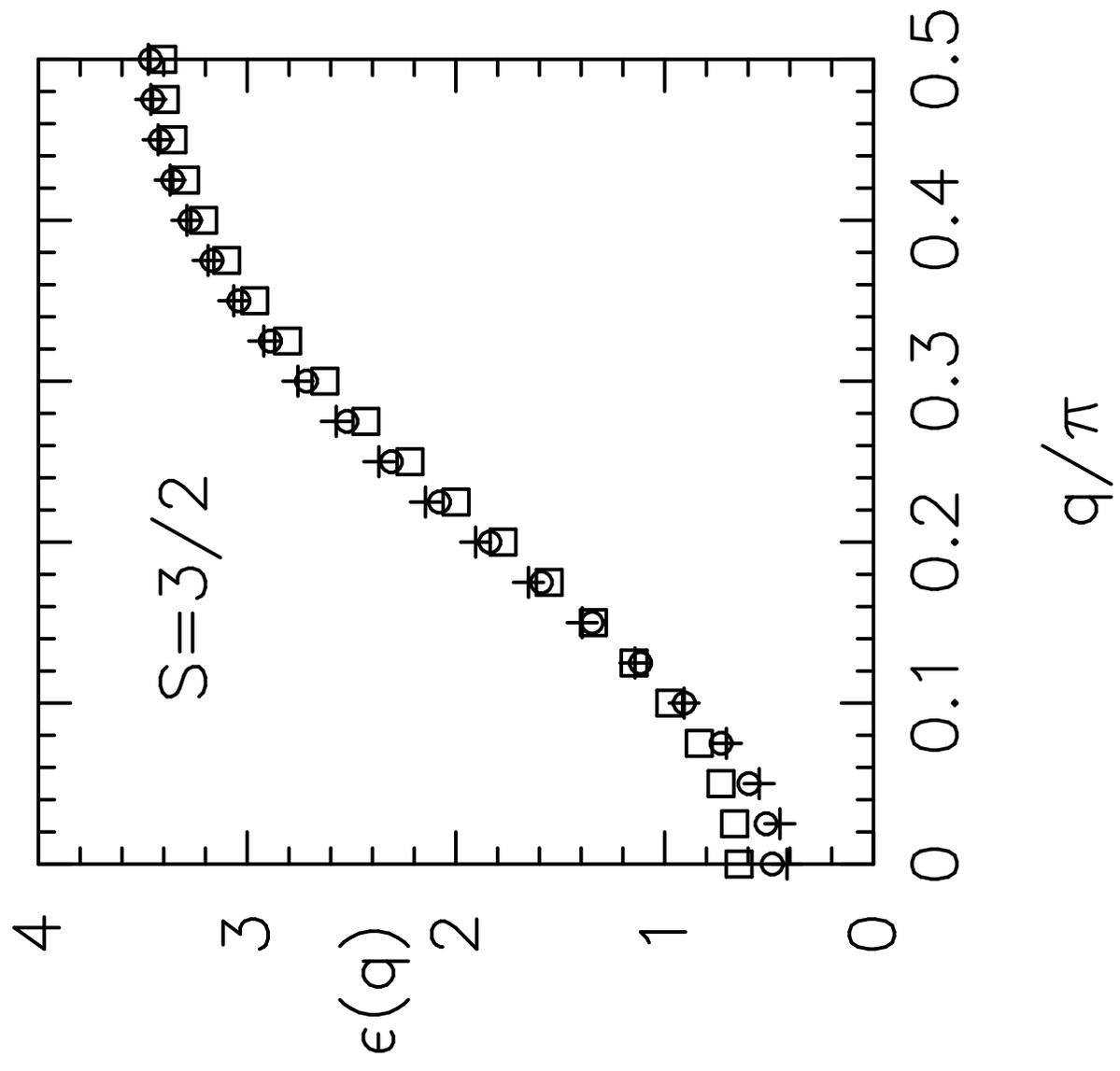
\end{document}